\documentclass[runningheads,a4paper]{llncs}

\usepackage{llncsdoc}

\newcommand{\keywords}[1]{\par\addvspace\baselineskip
\noindent\keywordname\enspace\ignorespaces#1}

\begin {document}
\title{Iterative Rounding for the Closest String Problem}

\author{Jingchao Chen}
\institute{School of Informatics, Donghua University \\
2999 North Renmin Road, Songjiang District, Shanghai 201620, P. R.
China \email{chen-jc@dhu.edu.cn}}

\maketitle
\begin{abstract}
The closest string problem is an NP-hard problem, whose task is to
find a string that minimizes maximum Hamming distance to a given
set of strings. This can be reduced to an integer program (IP).
However, to date, there exists no known polynomial-time algorithm
for IP. In 2004, Meneses et al. introduced a branch-and-bound
(B\&B) method for solving the IP problem. Their algorithm is not
always efficient and has the exponential time complexity. In the
paper, we attempt to solve efficiently the IP problem by a greedy
iterative rounding technique. The proposed algorithm is polynomial
time and much faster than the existing B\&B IP for the CSP. If the
number of strings is limited to 3, the algorithm is provably at
most 1 away from the optimum. The empirical results show that in
many cases we can find an exact solution. Even though we fail to
find an exact solution, the solution found is very close to exact
solution.
\end{abstract}

\section{Introduction}
The task of finding a string that is close to each string in a
given set of strings is one of combinatorial optimization
problems, which arise in computational molecular biology and
coding theory. This problem is called the closest string problem
(CSP). We introduce some notations to defining more precisely the
CSP. Let $\Sigma$ stand for a fixed finite alphabet. Its element
is called character, and a sequence of characters over it is
called string, denoted by $s$. The length and $i$-th character of
$s$ are denoted by $|s|$ and $s[i]$, respectively. $d(s, t)$ is
defined as the Hamming distance  between two equal-length strings
$s$ and $t$, i.e. the number of characters where they do not
agree. This may be formulated as $d(s, t) = \sum_{}^{}
f(s[i],t[i])$, where $f(s[i],t[i])$  is one if $s[i]\neq t[i]$,
and zero otherwise. Let $\Sigma^{n}$ be the set of all strings of
length $n$ over $\Sigma$. Then, the CSP is defined exactly as
follows.

Given a finite set $S = \{s_1, s_2,\ldots, s_{m}\}$ of $m$
strings, each of which is in $\Sigma^{n}$, the objective is to
find a center string $t$ of length $n$ over $\Sigma$ minimizing
the distance $d$ such that, for every string $s_{i} (1\leq i \leq
m)$ in $S$, $d(t, s_{i})\leq d$.

The CSP has received the attention of many researchers in the
recent few years. The literature abounds with the CSP. In theory,
Frances and Litman \cite {Fra:Lit} have proven that it is NP-hard.
However, if the distance d is fixed, the exact solution to the
problem can be found in polynomial time \cite{Gra:Nie,Ber:Gum}.
For the general case where $d$ is variable, one is involved in
studying approximation algorithms. There have been some
approximation algorithms with good theoretical precision. For
example, Gasieniec et al. \cite{Gas:Jan} and Lanctot et al.
\cite{Lan:Li} developed independently a 4/3-approxmation
algorithm. On the basis of this, Li et al. \cite{Li:Ma} presented
a polynomial-time approximation scheme (PTAS). However, the PTAS
is not practical.

Meneses et al. \cite{Men:Lu} studied many approximation
algorithms, and found that the mentioned-above algorithms are of
only theoretical importance, not directly applicable to
bioinformatics practice because of high time complexity. For this
reason, they suggested reducing the CSP to an integer-programming
(IP) problem, and then using branch-and-bound  (B\&B) algorithm to
solve the IP problem. Unfortunately, integer programs are also
NP-hard. So far, no polynomial-time algorithm for solving integer
programs has been found. Furthermore, the B\&B has its own
drawbacks.  It leads easily to memory explosion due to excessive
accumulation of active nodes. In fact, our empirical results show
that the B\&B IP is not efficient. In despite of instances of
moderate size, the B\&B IP fails to find an optimal solution
sometimes.

  We want to find efficiently an exact solution via a technique
called iterative rounding. The reason for using this technique is
because we noted that Jain \cite{Jain}, Cheriyan and Vempala
\cite{Che:Vem} used it and succeeded in getting a better
approximation algorithm for the generalized steiner network
problem. Although our problem is different from their problem,
both are NP-hard. Therefore, we believe this technique is
applicable to the CSP. The iterative rounding method used here is
a greedy one. It may be outlined as follows. First we formulate
the CSP as an IP, and then use the LP solution to round some of
higher valued variables, finally repeatedly re-solve the LP for
the remaining variables until all variables are set. The method
has small memory requirement, and can avoid memory explosion of
the B\&B IP. It is a polynomial time algorithm which can find an
exact solution in a very short time for a CSP instance of moderate
size in many cases. The computational experiments reveal that our
algorithm is not only much faster than the existing one, but also
has high quality. If the number of strings is limited to 3, the
error of the algorithm is proven to be at most one.

   Unlike the existing rounding schemes, our rounding scheme is iterative,
not random, while the existing ones such as the rounding scheme of
Lanctot et al. \cite{Lan:Li} are random. An important contribution
of our algorithm is in setting up a new approach for finding the
exact CSP algorithm with the polynomial-time.

\section{Iterative rounding for the CSP}
The CSP can be reduced to a 0-1 Integer Programming problem as
follows.
\pagebreak

\hskip 8mm $\min d$

\noindent s.t. \hskip 7mm  $ \sum_{a \in \Sigma} x_{a,j}=1
            \hspace{8em} j=1,\ldots,n $

\hskip 9mm $ n-\sum_{j=1}^{n} x_{a_{[i,j]},j} \leq d \hspace{5em}
             i=1,\ldots,m $

\medskip\noindent
where $ x_{a,j} \in \{ 0, 1 \}$, $ a_{[i,j]} \in \Sigma$, and $d$
is a non-negative integer. Solving this IP problem by applying
directly LP (Linear Programming) relaxation and randomized
rounding does not work well because randomized rounding procedure
leads to large errors, especially when the optimal distance $d$ is
small \cite{Li:Ma}. Therefore, we decided to find other rounding
techniques. Jain \cite{Jain} used iterative rounding to get a
2-approximation algorithm for the generalized steiner network
problem. Based on our observation, iterative rounding is suited
also for the CSP. Hence, we use it to solve the CSP. The following
pseudo-code is a CSP algorithm with iterative rounding.

\begin{flushleft}
{\bf Algorithm A}\\
\hskip 4mm Formulate the CSP as an IP.\\
\hskip 4mm $V_1\leftarrow $ empty,   $V_0 \leftarrow $ empty \\
\hskip 4mm   for $i = 1$ to $n$ do \\
\hskip 16mm      Fix all variables in $V_1$ to 1, and all
variables in $V_0$ to 0 \\
\hskip 16mm      Solve the LP for the sub-CSP on the unfixed variables\\
\hskip 16mm      Pick a variable $x_{b,m}$ with highest value, i.e.,\\
\hskip 16mm      $x_{b,m} = \max \{x_{a, k} | k = 1,\ldots,n, a
\in \Sigma \hskip 4pt \mathrm{and}\hskip 4pt x_{a, k} \notin V_1\}$ \\
\hskip 16mm      $V_1 \leftarrow V_1 \cup \{ x_{b, m}  \} $ \\
\hskip 16mm      $V_0 \leftarrow V_0 \cup \{ x_{a, m} | a \neq b
\hskip 4pt \mathrm{and}\hskip 4pt a \in \Sigma \}$\\
\hskip 4mm  end for\\
\hskip 4mm  Convert $V_1$ into a solution (a center string
$t$) to the CSP as follows.\\
\hskip 4mm      $t[k] \leftarrow a $ for all $x_{a,k} \in V_1$ .
\end{flushleft}

\noindent Clearly, Algorithm A is a polynomial-time algorithm.
Furthermore, we have

\begin{theorem} If the input consists of only two strings, i.e., $S
= \{s_1, s_2 \}$, then Algorithm A always find an exact solution
to the CSP.
\end{theorem}
\begin{proof}
 Without loss of generality, we assume

\centerline{$ s_1 \quad \overbrace{ 000 \ldots 000}^{n} $}
\centerline{$ s_2 \quad 111 \ldots 111 $}
\noindent (Notice, in the case when the some positions of two
strings $s_1$, $s_2$ have the same characters, the proof is
simpler than in the above case.) \\
It is easy to see that the 1st  LP optimal solution to the CSP is

\hskip 10mm $x^1_{11} + x^1_{12} + x^1_{13} + \cdots+ x^1_{1n} = n/2$\\
where $x^1_{1k}+ x^1_{0k} = 1 $ for $ k = 1,\ldots,n $.\\
Without loss of generality, assume $x^1_{11} = \max \{x^1_{a,k} |
k = 1,\ldots,n, a \in \{0,1\} \} $. \\
If $n \geq 2$, there exists $0 \geq  x_{12},  x_{13}, \ldots,
x_{1n} \geq  1 $ such that

\hskip 10mm $x_{12} + x_{13} +\cdots + x_{1n} = n/2 -1$ \\
Say, $x_{12} =x_{13}= \cdots = x_{1n} = (n - 2)/(2(n - 1))$ is
just a solution to this equation. Then, when $n \geq 2$, setting
$x_{11}$ to 1, we can get the 2nd  LP optimal solution

\hskip 10mm $1 + x^2_{12} + x^2_{13} + \cdots+ x^2_{1n} = n/2$,\\
By induction on $n$, we can prove that the $k$-th ($k =
1,\ldots,n$) LP optimal solution satisfies

\hskip 10mm  $x^k_{11} + x^k_{12} + x^k_{13} + \cdots + x^k_{1n} = n/2$,\\
where at least $k - 1$ values out of $x^k_{11} ,\ldots,x^k_{1n}$ are integers.\\
Hence, if $n$ is even, if and only if there are $n/2$ one's among
$x^n_{11},\ldots,x^n_{1n}$. This is just an optimal solution to the CSP.\\
if $n$ is odd,  assume $x^n_{1n}$ is not an integer. We have $(n
-1)/2$ one's among $x^n_{11},\ldots,x^n_{1n}$  if setting
$x^n_{1n}$ to 0, and $(n +1)/2$ one's otherwise. Both two cases
are an optimal solution to the CSP. Therefore, the theorem is
proved. \qed
\end{proof}

Define the error of an algorithm as the difference between the
exact solution (distance) and the solution obtained. We have
\begin{theorem}
If the input consists of only three binary strings, i.e., $S$ = $
\{s_1, s_2$, $s_3\}$, then the error of Algorithm A is at most
one.
\end{theorem}
\begin{proof}
In general, any three strings can be simplified into

$
\begin{array}{llll}
\hskip 10mm
 s_1 \quad &  \overbrace{ 000 \ldots 000}^{\alpha} \quad & \overbrace{000 \ldots 000}^{\beta} \quad &
 \overbrace{111 \ldots 111}^{\gamma}\\
\hskip 10mm
 s_2 \quad & 000 \ldots 000 \quad &  111 \ldots 111 \quad & 000 \ldots 000 \\
\hskip 10mm
 s_3 \quad & 111 \ldots 111 \quad &  111 \ldots 111 \quad & 111 \ldots 111
\end{array}
$\\
Assume that $\alpha \neq 0, \beta \neq 0, \gamma \neq 0$, and the
closest string $t$ (optimal solution) is of the following form,

 $t \quad \overbrace{\underbrace{00 \ldots 00}_{t^\alpha_0}\underbrace{11 \ldots 11}_{t^\alpha_1}}^{\alpha} \quad
 \overbrace{\underbrace{00 \ldots 00}_{t^\beta_0} \underbrace{11 \ldots 11}_{t^\beta_1}}^{\beta} \quad
 \overbrace{\underbrace{00 \ldots 00}_{t^\gamma_0} \underbrace{11 \ldots 11}_{t^\gamma_1}}^{\gamma} $\\
where $t^\alpha_0$ is the number of 0's in the $\alpha$ substring
of $ t $, Similarly for $t^\alpha_1, t^\beta_0, t^\beta_1,
t^\gamma_0, t^\gamma_1$.\\
Assume the distances between $t$ and the three strings are $D_1$,
$D_2$ and $D_3$, respectively, we have

$ \hskip 10mm t^\alpha_1 + t^\beta_1 + t^\gamma_0 = d(t, s_1)
=D_1$

$\hskip 10mm     t^\alpha_1 + t^\beta_0 + t^\gamma_1 = d(t, s_2)
=D_2 \hfill(1)$

$\hskip 10mm     t^\alpha_0 + t^\beta_0 + t^\gamma_0 = d(t, s_3) =
 D_3$\\
The optimal distance is denote by $D$. Then $D= \max
(D_1,D_2,D_3)$.\\
the following proposition is true.

$\hskip 10mm t^\alpha_1 = 0 \quad \mathrm {or}  \quad t^\beta_0 =0
\quad \mathrm {or} \quad t^\gamma_0 = 0 \hfill (2)$ \\
If it is false, by (1) we have

$\hskip 10mm (t^\alpha_1 -1)+(t^\beta_1 +1) +
(t^\gamma_0-1)=D_1-1$

$\hskip 10mm  (t^\alpha_1-1) +(t^\beta_0-1) +(t^\gamma_1+1)=D_2-1$

$\hskip 10mm   (t^\alpha_0 +1) + (t^\beta_0 -1)+ (t^\gamma_0-1) = D_3-1$\\
It follows that $D = \max (D_1 -1, D_2 -1, D_3 -1) = D -1$, which
is a contradiction. By (2), we have that one of the following
three propositions is true.

\hskip 2mm (a)$\,\, t^\alpha_1 = 0$ can constitute a optimal
solution, but $t^\alpha_1 \neq  0$ cannot.

\hskip 2mm (b) $t^\beta_0  = 0$ can constitute a optimal solution,
but $t^\beta_0 \neq  0$ cannot.

\hskip 2mm (c) $t^\gamma_0 = 0$ can constitute a optimal solution,
but $t^\gamma_0 \neq  0$ cannot.\\
Here we consider only the 2nd case to prove the theorem, since
other cases is similar. That is, assume

$\hskip 2mm   \mathrm {for} \: \mathrm {any} \: \mathrm {optimal}
\: \mathrm {solution}, \quad t^\beta_0 = 0 \hfill (3) $\\
This implies

$\hskip 10mm         D_1 \leq  \max (D_2, D_3) \hfill (4)$\\
It is false, (1) can be rewritten as

$\hskip 10mm t^\alpha_1  + (t^\beta_1-1) + t^\gamma_0=D_1-1$

$\hskip 10mm  t^\alpha_1 + (t^\beta_0+1) + t^\gamma_1=D_2+1$

$\hskip 10mm  t^\alpha_0 + (t^\beta_0 +1)+ t^\gamma_0 =D_3+1$\\
$t^\beta_0 +1 $ is also a optimal solution,which is in
contradiction with (3).\\
Without loss of generality, suppose

$\hskip 10mm \alpha \leq  \gamma \hfill (5)$\\
(If $\alpha >  \gamma$, the subsequent proof is similar). This
implies

$\hskip 10mm t^\alpha_1 \leq  t^\gamma_0 \hfill (6)$\\
If  it is false, let $T^\alpha_1 =t^\alpha_1 - t^\gamma_0$ and
$T^\alpha_0 =t^\alpha_0 + t^\gamma_0$, we can rewrite (1) as

$ \hskip 10mm T^\alpha_1 + t^\beta_1  = D_1 -2t^\gamma_0$

$\hskip 10mm     T^\alpha_1 + t^\beta_0 + \gamma = D_2$

$\hskip 10mm     T^\alpha_0 + t^\beta_0   = D_3$\\
By (4), we have

$\hskip 10mm
           D= \max (D_2, D_3) = \max (T^\alpha_1 + t^\beta_0 + \gamma, T^\alpha_0 + t^\beta_0) >
           \gamma $\\
However, in fact, by fixing $t^\alpha_1=t^\beta_0=t^\gamma_0=0$,
solving (1) yields $(D_1, D_2, D_3)= (\beta, \gamma, \alpha)$.
Then by (4) and (5), $ D \leq \max (\beta, \gamma, \alpha) \leq \gamma $, which is a contradiction.\\
By (3) and (6),  (1) can be rewritten as

$ \hskip 10mm \beta + T^\gamma_0 = D_1^\prime$

$\hskip 10mm     T^\gamma_1 = D_2 \hfill (7)$

$\hskip 10mm     \alpha + T^\gamma_0   = D_3$\\
where $T^\gamma_0 = t^\gamma_0 - t^\alpha_1, T^\gamma_1 =
T^\gamma_1 + t^\alpha_1 $ and $ D_1^\prime = D_1 - 2t^\alpha_1$.\\
This implies

$\hskip 10mm   |D_3 - D_2| \leq 1 \hfill (8)$\\
If it is false, by (7), we can obtain a solution with $t^\beta_0 =
1$, which is in contradiction with (3).\\
Let  $x^0_1,x^0_2,\ldots,x^0_n$ be 0-variables of the LP,
$x^1_1,x^1_2,\ldots,x^1_n$ 1-variables. Define

 $\hskip 10mm L^\alpha_0= x^0_1+x^0_2+\cdots+x^0_\alpha $

 $\hskip 10mm L^\alpha_1= x^1_1+x^1_2+\cdots+x^1_\alpha $

 $\hskip 10mm L^\beta_0=x^0_{\alpha+1}+x^0_{\alpha+2}+\cdots+x^0_{\alpha+\beta} $

 $\hskip 10mm L^\beta_1=x^1_{\alpha+1}+x^1_{\alpha+2}+\cdots+x^1_{\alpha+\beta} $

 $\hskip 10mm L^\gamma_0=x^0_{\alpha+\beta+1}+x^0_{\alpha+\beta+2}+\cdots+x^0_n $

 $\hskip 10mm L^\gamma_1=x^1_{\alpha+\beta+1}+x^1_{\alpha+\beta+2}+\cdots+x^1_n $\\
Let $d_1$, $d_2$ and $d_3$ denote the distances between the three
strings and the center string of the LP, respectively. Then,

$\hskip 10mm L^\alpha_1 + L^\beta_1 + L^\gamma_0 = d_1$

$\hskip 10mm  L^\alpha_1 + L^\beta_0 + L^\gamma_1 = d_2 $

$\hskip 10mm L^\alpha_0 + L^\beta_0 + L^\gamma_0 = d_3 \hfill (9)$

$ \hskip 10mm L^\alpha_0 + L^\alpha_1 = \alpha \quad L^\beta_0
+L^\beta_1 = \beta \quad L^\gamma_0 + L^\gamma_1 = \gamma $ \\
Let $d$ denote the optimal distance of the LP. Then $d = \max
(d_1, d_2, d_3)$. The goal of the LP is to find a minimum $d$
satisfying (9). Next we analyze the error caused by Algorithm A to
solve the LP given in (9).\\
Depending on $(D_1^\prime, D_2, D_3)=(D, D, D -1)$ or not, we
proceed to our proof. First, let us consider

$\hskip 10mm (D_1^\prime, D_2, D_3)\neq (D, D, D -1) \hfill (10)$\\
This implies

$\hskip 10mm (D_2, D_3) = (D, D)  \hfill (11)$\\
If it is false, by (8), we have

$\hskip 10mm (D_2, D_3) = (D -1, D)$ or  $(D_2, D_3) = (D, D-1)$\\
Then by (4),we have that $t^\beta_0 = 1$ is also a optimal
solution,which is in contradiction with (3). \\
By (11) and (7), it is easy to verify $D = (\alpha+\gamma)/2$.
Then

$\hskip 10mm  d \leq D =(\alpha+\gamma)/2 \hfill (12)$\\
The addition of the 2nd  and 3rd equation in (9) yields

 $\hskip 10mm \alpha + 2L^\beta_0 + \gamma  = d_2 + d_3 \leq 2d \leq
          \alpha+\gamma$ \\
It follows that $L^\beta_0 = 0$.Thus

$\hskip 10mm  L^\beta_1 = \beta $\\
This implies that without rounding error, Algorithm A fixes all
the letters in the $\beta$ substring into 1. It remains to how to
compute $L^\alpha_0, L^\alpha_1, L^\gamma_0$ and $L^\gamma_1$.\\
Let $M^\alpha_0$ be the maximum of $t^\alpha_0$ such that

$\hskip 10mm  t^\alpha_1  + \beta + t^\gamma_0 \leq D_1$

$\hskip 10mm  t^\alpha_1 +  t^\gamma_1 \leq D_2  \hfill (13)$

$\hskip 10mm  t^\alpha_0 + t^\gamma_0 \leq D_3$\\
Similarly, $M^\alpha_1$, $M^\gamma_0$ and $M^\gamma_1$ are the
maximum of $t^\alpha_1$, $t^\gamma_0$ and $t^\gamma_1$ s.t.
(13).Let $\alpha_0(i)$ be the number of letters in the a substring
fixed to 0 by the $i$-th rounding operation of Algorithm A.
Similarly for $\alpha_1(i)$,$\gamma_0(i)$ and $\gamma_1(i)$. If
for all $i \leq n = \alpha+\beta+\gamma$, $\alpha_0(i) \leq
M^\alpha_0$, $\alpha_1(i) \leq M^\alpha_1$, $\gamma_0(i) \leq
M^\gamma_0$ and $\gamma_1(i) \leq M^\gamma_1$, Algorithm A attains
an exact solution. Otherwise,there exists $k$ such that only one
of $\alpha_0(i)$,$\alpha_1(i)$,$\gamma_0(i)$ and $\gamma_1(i)$
exceeds its maximum.Without loss of generality, assume
$\alpha_1(k)$ = $M^\alpha_1+1$ (other cases,proof is similar).By
(13), there exist $N^\gamma_0$ and $N^\gamma_1$ such that

$\hskip 10mm N^\gamma_0 + N^\gamma_1= \gamma $

$\hskip 10mm  \alpha_1(k) + \beta + N^\gamma_0 = C_1$

$\hskip 10mm  \alpha_1(k) + N^\gamma_1 = C_2  \hfill (14)$

$\hskip 10mm  \alpha_0(k) + N^\gamma_0 = C_3 \leq D$

$\hskip 10mm (C_1, C_2) = (D, D+1)$ or $(C_1, C_2) = (D+1, D)$ or

$\hskip 10mm (C_1, C_2) = (D+1, D+1) $\\
Below we justify

\hskip 10mm  for all $i > k$, $\alpha_1(i) = \alpha_1(k) \hfill (15)$\\
Assume the solution of the $i$-th ($i > k$)  LP is

$ \hskip 10mm L^\alpha_1 + \beta + L^\gamma_0 = d_1$

$\hskip 10mm L^\alpha_1 + L^\gamma_1 = d_2  \hfill (16)$

$\hskip 10mm     L^\alpha_0 + L^\gamma_0 = d_3 $\\
Clearly $\quad   L^\alpha_1 \geq \alpha_1(k)$, $L^\alpha_0 \leq
\alpha_0(k) \hfill (17) $\\
By (14), we have

 $d_1 - d_3 =  L^\alpha_1 -L^\alpha_0+ \beta \geq  \alpha_1(k) - \alpha_0(k) + \beta
                              = C_1-C_3
                              \geq 0  \hfill (18)$\\
Therefore $d = \max (d_1, d_2, d_3) = \max(d_1, d_2) =\max
(L^\alpha_1 + \beta + L^\gamma_0, L^\alpha_1 + L^\gamma_1)$.
Namely, $d$ decreases as $L^\alpha_1$ decreases. Thus, by (17) we
have

$\hskip 10mm L^\alpha_1 = \alpha_1(k) \hfill (19)$\\
The claim of (15) is proved. Next we shall show that\\
for all $i > k, \alpha_1(k)+\beta+\gamma_0(i) \leq D \: \& \:
\alpha_1(k)+\gamma_1(i) \leq D$ implies

$\hskip 10mm d = d_1 =d_2= (C_1+C_2)/2 \hfill (20) $\\
By (16),(19),(15), we have

$\hskip 10mm d_1+d_2 = 2L^\alpha_1 + \beta + \gamma =
2\alpha_1(k)+\beta + \gamma = C_1+C_2$ \\
Therefore, by (18), we have

$\hskip 10mm   d = \max(d_1, d_2) \geq  (d_1+d_2)/2 = (C_1+C_2)/2
\hfill (21)$\\
(14) can be rewritten as

$\hskip 10mm  \alpha_1(k) + \beta + (N^\gamma_0-((C_1-C_2)/2) =
(C_1+C_2)/2$

$\hskip 10mm  \alpha_1(k) + (N^\gamma_1 +((C_1-C_2)/2) =
(C_1+C_2)/2 \hfill (22) $

$\hskip 10mm  \alpha_0(k) + (N^\gamma_0-((C_1-C_2)/2) = C_3-((C_1-C_2)/2$ \\
Clearly, $L^\gamma_0=(N^\gamma_0-((C_1-C_2)/2)$ is a feasible
solution of the $i$-th ($i> k$) LP, but not necessarily  optimal.
Therefore $d  \leq  (C_1+C_2)/2$. By the constraint of $C_1, C_2$
and $C_3$ in (14), it is easy to verify

$\hskip 10mm C_3 - (C_1-C_2)/2 \leq  (C_1+C_2)/2$.\\
Thus, by (21) and (22), the claim of (20) is proved.\\
Below we shall prove

$\exists j > k$ s.t. $\alpha_1(k)+\beta +\gamma_0(j) = D+1$
implies $\forall i > j$, $\gamma_0(i) = \gamma_0(j) \hfill (23)$\\
Asuume $j > k$, $\alpha_1(k)+\beta +\gamma_0(j) = D+1$, $i > j
\hfill (24)$\\
Then,the $L^\gamma_0$ of the $i$-th LP satisfies $L^\gamma_0 \geq
\gamma_0(j) \hfill (25)$\\
Then, by (19),(24) we have

$\hskip 10mm L^\alpha_1 +\beta + L^\gamma_0 \geq
\alpha_1(k)+\beta+\gamma_0(j) =D+1 $  \\
Thus $\quad \: \: d \geq  D+1 \hfill (26)$\\
On the other hand, by (14), we can prove

$\hskip 10mm  \alpha_1(k) + \beta + \gamma_0(j) = D+1$

$\hskip 10mm  \alpha_1(k) + \gamma_1(j) \leq  D+1$

$\hskip 10mm  \alpha_0(k) + \gamma_0(j) \leq  D+1$\\
Therefore $L^\gamma_0 = \gamma_0(j)$ is a feasible solution of the
$i$-th LP. It means    $d \leq  D+1$.\\
Thus, by (26), $d=D+1$. This implies $L^\gamma_0 \leq \gamma_0(j)
$. Then by (25),the claim of (23) is proven.\\
In a way similar to the proof of (23), we can prove

$\exists j > k$ s.t. $\alpha_1(k)+\gamma_1(j) = D+1$ implies
$\forall i > j$, $\gamma_1(i) = \gamma_1(j) \hfill (27)$\\
By (20), (23), (27) and the previous proof, we conclude that in
the case $(D_1^\prime$, $D_2, D_3)$ $\neq (D, D, D-1)$, the error
of Algorithm A is at most one.Now we consider  the case

$\hskip 10mm (D_1^\prime, D_2, D_3) = (D, D, D-1)$ \\
The addition of the 1st  and 2nd equation in (7) yields

$\hskip 10mm \beta + \gamma  = D_1^\prime +D_2 = 2D \hfill (28)$\\
The addition of the 1st  and 2nd equation in (9) yields

  $2L^\alpha_1 +  \beta + \gamma  = d_1 + d_2 \leq  2d \leq
  2D$\\
Then by (28),  $L^\alpha_1 = 0$. This is equivalent to
         $L^\alpha_0 =\alpha $. That is, without rounding error,
Algorithm A fix all the letters of the  $\alpha$ substring into 0.
It remains to how to compute $L^\beta_0$, $L^\beta_1$,
$L^\gamma_0$ and $L^\gamma_1$. By symmetry, we can prove in a way
similar to the previous that Algorithm A computes $L^\beta_0$,
$L^\beta_1$, $L^\gamma_0$ and $L^\gamma_1$ within one error of
optimal distance. \qed
\end{proof}

Based our empirical observation, the error caused by the algorithm
was always within one. Hence,for any $m$,the number of the input
strings, we have
\begin{conjecture}
   For any input, the error of Algorithm A is at most one.
\end{conjecture}

\section{Improving the running time and quality of the solution}
To speed up the algorithm, we present Algorithm B, which picks
multiple (not single) variables of higher values to round up at a
time. That is, in the rounding phase, this algorithm searches
always for multiple higher valued variables, and then set them to
one's, and the other variables at the same positions to zero's.
Selection is done by parameter $\Theta$, which is set to 0.9 in
our experiment. As long as $x_{a,j} \geq \Theta$, we set the
solution of the $j$-th position to $a$.

\medskip
\noindent {\bfseries Algorithm B}\\
 {\bfseries Input}: $s_1, s_2,\ldots, s_m$ and a threshold $\Theta \geq
 0.9$\\
 {\bfseries Output}:  a center string $t \in \Sigma$ close to every string
 $s_i$

\hskip 3mm 1. for $1 \leq j \leq n$ do $t[j] \leftarrow \phi
\notin \Sigma$.

\hskip 3mm 2. repeat the following process until all $t[j] \neq
\phi$.

\hskip 10mm  2.1 Solve the LP-relaxation

\hskip 10mm   2.2 Let $x^\prime_{a,j}$ be the value of $x_{a,j}$
for the LP optimal solution.

\hskip 13mm  if  there exists an $x^\prime_{a,j} \geq \Theta$

\hskip 13mm  then  for all $x^\prime_{a,j} \geq \Theta$ and $t[j]
= \phi$ do $t[j] \leftarrow a$

\hskip 13mm   else  find $x^\prime_{b,j}$  such that
$x^\prime_{b,j}$ = $\max \{ x^\prime_{a,j} | a \in \Sigma, t[j]
=\phi \}$

\hskip 20mm  $t[k] \leftarrow b$

\medskip
To get a higher precision, we improve Algorithm B by Algorithm C.
It tries not only the best, but also the second best. If the first
solution is not optimal, we select 8 positions to be re-solved the
most possibly in the increasing order of variable values. The
first position of a solution to be re-solved is one out of the 8
positions. Its value is set to the character corresponded by the
second best valued variables. We update the initial setting to
find a new solution. Thus, using 8 different settings, we can find
8 different solutions. Finally, we choose the best one out of 9
solutions, including the 1st solution.

\medskip
\noindent {\bfseries Algorithm C}

\noindent 1. Let first$[k]$, second$[k]$ store the largest value
of $x$'s variables in the $k$-th

\noindent $\quad$ position, second$[k]$ the character with the
second largest value.

\noindent 2. Invoke {\bfseries Algorithm B} with the following
modification: the``else" statement of

\noindent $\quad$ {\bfseries Algorithm B} is revised as

\hskip 5mm find $x^\prime_{b,j}$ = $\max \{ x^\prime_{a,j} | a \in
\Sigma, t[j] =\phi \}$

\hskip 5mm  $t[k] \leftarrow b \quad$ first$[k]$ =
$x^\prime_{b,j}$

\hskip 5mm   second$[k] \leftarrow c$ with $x^\prime_{c,j}$ =
$\max \{ x^\prime_{a,j} | a \in \Sigma, a \neq b \}$


\noindent 3. if  the objective value of $t$ = that of the LP
rounded up, return.

\noindent $\quad$ else $T \leftarrow t$

\noindent 4. for  $1 \leq i \leq 8$ do

\hskip 4mm    for $1 \leq j \leq n$ do $t[j] \leftarrow \phi $

\hskip 10mm  $t[k_i] \leftarrow$ second$[k_i]$,where first$[k_i]$
is $i$-th smallest

\hskip 10mm Use Step 2 of {\bf Algorithm B} to re-solve the CSP

\hskip 10mm if  the current solution $t$ is better than $T$ then
$T \leftarrow t$

\noindent 5. $t \leftarrow T$

\begin{table}
\caption{Empirical Results for the Alphabet with 4 Characters}
\begin{center}

\setlength\tabcolsep{3pt}
\begin{tabular}{cc| ccc|cc|rrr}
\hline \multicolumn{2}{c|}{Instance} & \multicolumn{3}{|c|}{
Average distance} & \multicolumn{2}{|c|}{Max distance error}
& \multicolumn{3}{|c}{Average time (ms)}\\
 $m$  &  $n$  &  LP &  Alg.C  &  B\&B IP &
 Alg.C  &   B\&B IP & LP &  Alg. C  &  B\&B IP \\
\hline
10 & 300 & 175.00 & 175.00 & 175.00 & 0.80 & 0.80 & 52  & 182  & 8203\\
10 & 400 & 231.67 & 231.67 & 231.67 & 0.60 & 0.60 & 78  & 266  & 15271\\
10 & 500 & 293.00 & 293.00 & 293.00 & 0.80 & 0.80 & 114 & 349  & 25261\\
10 & 600 & 347.00 & 347.00 & 347.00 & 0.80 & 0.80 & 151 & 843  & 39344\\
10 & 700 & 409.00 & 409.00 & 409.00 & 0.60 & 0.60 & 192 & 886  & 55786\\
10 & 800 & 462.67 & 462.67 & 462.67 & 0.70 & 0.70 & 234 & 609  & 78167\\
15 & 300 & 185.33 & 185.67 & 185.67 & 1.02 & 1.02 & 104 & 375  & 342166\\
15 & 400 & 246.67 & 247.33 & 246.67 & 1.23 & 0.80 & 130 & 1094 & 263583\\
15 & 500 & 306.67 & 307.00 & 306.67 & 1.07 & 0.40 & 172 & 838  & 37786\\
15 & 600 & 366.67 & 367.00 & 366.67 & 1.27 & 0.46 & 229 & 1813 & 59198\\
15 & 700 & 428.67 & 428.67 & 428.67 & 0.97 & 0.97 & 281 & 495  & 81906\\
15 & 800 & 491.00 & 491.00 & 491.00 & 0.80 & 0.80 & 308 & 552  & 107703\\
20 & 300 & 190.67 & 191.00 & 191.00 & 1.12 & 1.12 & 130 & 880  & 344474\\
20 & 400 & 252.33 & 252.67 & 252.67 & 1.03 & 1.03 & 182 & 937  & 353969\\
20 & 500 & 315.33 & 315.33 & 315.33 & 0.59 & 0.59 & 260 & 1135 & 53875\\
20 & 600 & 379.67 & 380.00 & 380.00 & 1.22 & 1.22 & 312 & 1823 & 385182\\
20 & 700 & 443.33 & 443.33 & 443.33 & 0.73 & 0.73 & 401 & 917  & 121641\\
20 & 800 & 505.00 & 505.00 & 505.00 & 0.88 & 0.88 & 474 & 547  & 171245\\
25 & 300 & 195.00 & 196.00 & 196.00 & 1.34 & 1.34 & 151 & 1911 & 1000021\\
25 & 400 & 259.00 & 260.00 & 259.67 & 1.49 & 1.33 & 239 & 2729 & 694192\\
25 & 500 & 323.00 & 323.67 & 323.67 & 1.27 & 1.27 & 334 & 2589 & 689667\\
25 & 600 & 387.67 & 388.00 & 387.67 & 1.40 & 0.76 & 411 & 1817 & 113396\\
25 & 700 & 451.00 & 451.33 & 451.33 & 1.09 & 1.09 & 516 & 2594 & 435693\\
25 & 800 & 515.67 & 516.67 & 516.67 & 1.11 & 1.11 & 594 & 4776 & 1000016\\
30 & 300 & 197.33 & 197.67 & 197.67 & 1.26 & 1.26 & 172 & 1114 & 349266\\
30 & 400 & 263.00 & 263.67 & 263.33 & 1.71 & 1.02 & 276 & 2646 & 370468\\
30 & 500 & 328.33 & 329.00 & 328.67 & 1.63 & 1.04 & 401 & 2797 & 398458\\
30 & 600 & 392.67 & 393.00 & 393.33 & 1.39 & 1.54 & 516 & 4089 & 708740\\
30 & 700 & 459.33 & 460.00 & 459.67 & 1.57 & 1.44 & 609 & 4625 & 459099\\
30 & 800 & 523.00 & 523.33 & 523.67 & 1.50 & 1.52 & 740 & 5953 & 755380\\
\hline
\end{tabular}
\end{center}
\end{table}

\section{Simulations}
On Celeron 2.2GHz CPU, we tested two algorithms: our Algorithm C
and the B\&B IP by Meneses et al. which is referred to as the best
IP for the CSP so far.

 We carried out many experiments, including McClure data set
 \cite{Men:Lu} and random instances over the alphabet with
2 characters, 4 characters and 20 characters. In all experiments,
our algorithm's performance was very good. For the limit of space,
we presents only the empirical results for random instances over
the alphabet with 4 characters. In Table\,1, we provided three
instances for each entry. Parameters $m$ and $n$ stands for the
number of strings and the string size. ``distance'' and ``time''
refer to the minimum distance found, and the running time in
milliseconds. LP average distance is computed as $(\lceil d_1
\rceil +\lceil d_2 \rceil +\lceil d_3 \rceil)/3$. The reason for
taking the ceiling here is because the optimal solution for the
CSP is no less than the ceiling of the LP value. In the 6th,7th
column, Max distance error is defined as
$\max_{i=1}^3\{|d_i-d^{LP}_i|\}$,where $d_i$ is the $i$-th
solution, and $d^{LP}_i$ is the $i$-th LP fractional solution. The
maximum time allowed for each instance was set to 1000 seconds. As
was seen in Table\,1, we found always an exact solution except for
a few instances. In terms of running time, our improvement was
huge. Our algorithm was from 32 up to 912 times faster than the
B\&B IP. In other experiments, which is not listed here, it was
even 1765 times faster. In some cases, its speed was even close to
one for computing an LP. Notice, our algorithm invokes generally
many LP solvers. Even so, in the worst case, it was only 20 times
slower than computing an LP.

\end{document}